\documentclass[abstracton]{scrartcl}
\usepackage[utf8]{inputenc}

\usepackage[T1]{fontenc}
\usepackage{lmodern}
\usepackage{amsmath}
\usepackage[hidelinks]{hyperref}
\usepackage[noabbrev]{cleveref}
\usepackage{microtype}
\usepackage[exponent-product = \cdot]{siunitx}
\usepackage[affil-it]{authblk}

\usepackage{tikz}
\usepackage{pgfplots}
\pgfplotsset{compat=newest}
\usepackage[european, straightvoltages]{circuitikz}

\usepackage[backend = biber,
    style = numeric,
    giveninits = true,
    natbib = true,
    hyperref = true,
    maxbibnames = 10,
    sorting = none
    ]{biblatex}

\setkomafont{author}{\normalsize}
\setkomafont{date}{\normalsize}

\newcommand{\uix}[1]{_\mathrm{#1}}
\DeclareMathOperator{\csch}{csch}
\renewcommand{\j}{\mathrm{j}}

\definecolor{color1}{HTML}{FF7300}
\definecolor{color2}{HTML}{1570CC}
\definecolor{color3}{HTML}{D90100}
\definecolor{color4}{HTML}{07BB03}
\definecolor{color5}{HTML}{7E35CC}

\begin{document}

\title{Estimation of acoustic wave non-linearity in ultrasonic measurement systems}

\author{Leander Claes}
\author{Carolin Steidl}
\author{Tim Hetkämper}
\author{Bernd Henning}
\affil{Measurement Engineering Group, Paderborn University,\\ Warburger Straße 100, 33098 Paderborn, Germany}

\maketitle

\begin{abstract}
Most measurement methods based on ultrasound, such as sound velocity, absorption or flow measurement systems, require that the acoustic wave propagation is linear.
In many cases, linear wave propagation is assumed due to small signal amplitudes or verified, for example, by analysing the received signal spectra for the generation of harmonic frequency components.
In this contribution, we present an approach to quantify occurrence of non-linear effects of acoustic wave propagation in ultrasonic measurement systems based on the evaluation of the acoustic Reynolds number.
One parameter required for the determination of the acoustic Reynolds number is the particle velocity of the acoustic wave, which is not trivially obtained in most measurement systems.
We thus present a model-based approach to estimate the particle velocity of an acoustic wave by identifying a Mason model from electrical impedance measurements of a given transducer.
The Mason model is then used to determine the transducer's velocity output for a given electrical signal, allowing for an evaluation of the acoustic Reynolds number for different target media.
\end{abstract}

\section{Motivation}

Due to the character of the differential equations that describe the behaviour of fluids, such as the Navier-Stokes-Equation and the equation of state for the respective fluid, all acoustic wave propagation is non-linear.
When deriving the differential equation for an acoustic wave, one assumes the amplitude of the acoustic wave to be sufficiently small, so that the non-linear terms that exist in constituting equations are negligible~\cite{Landau2013}.
Thus, for applications of acoustic waves that are assumed to be linear, it has to be determined if the amplitude of the acoustic wave created by a given transducer satisfies the aforementioned condition.
One option to verify if the acoustic signal's particle velocity amplitude \(v_0\) is sufficiently small is to consider the acoustic Reynolds number \(N\uix{Re}\)~\cite{Rudenko1977}:
\begin{equation}
	N\uix{Re} = \frac{v_0 c \rho}{\mu \omega} ,
	\label{equ:acoustic_reynolds}
\end{equation}
where \(c\) denotes the sound velocity and \(\rho \) denotes the density of the medium.
\(\omega \) is the angular frequency of the acoustic wave and \(\mu \) are the combined linear thermal and viscous losses in the fluid:
\begin{equation}
	\mu = \frac{4}{3} \mu\uix{s} + \mu\uix{v} + \frac{c\uix{p} - c\uix{v}}{c\uix{p} \cdot c\uix{v}} \nu ,
\end{equation}
with the shear viscosity \(\mu\uix{s}\) and the volume viscosity \(\mu\uix{v}\).
\(c\uix{p}\) and \(c\uix{v}\) are the isobaric and isochoric specific heat capacities and \(\nu \) is the thermal conductivity of the fluid. 
The acoustic Reynolds number \(N\uix{Re}\) describes the relation of the accumulation of non-linear effects to the effects of linear absorption caused by the thermal and viscous losses \(\mu \).
Thus, for \(N\uix{Re} \gg 1\) non-linear effects are predominant, while for \(N\uix{Re} \ll 1\), linear absorption dominates the properties of acoustic wave propagation~\cite{Rudenko1977}.
For practical applications this relationship implies that the sound propagation tends to be more linear if the losses \(\mu \) in the medium are high.
This can be expressed as a lower bound for \(\mu \) if the requirement \(N\uix{Re} \ll 1\) is to be satisfied:  
\begin{equation}
	\mu \gg \frac{v_0 c \rho}{\omega} =  \frac{v_0 Z}{\omega}.
	\label{equ:losses_bound}
\end{equation}
In \cref{equ:acoustic_reynolds} and \cref{equ:losses_bound} the product of sound velocity \(c\) and density \(\rho \) can be replaced by the specific acoustic impedance of the medium \(Z\).

While the values of the sound velocity and the density of the medium, as well as the angular frequency \(\omega \) of the acoustic wave, are usually available and the losses \(\mu \) can be estimated, the particle velocity amplitude \(v_0\) is not trivially obtainable.
As the acoustic wave amplitude in a simplified acoustic measurement system usually has its maximum at the transmitting transducer's active surface, one can assume \(v_0\) to be the normal velocity of said surface.
While this velocity can be measured by means of laser vibrometry, the surface has to be loaded with the target medium during the measurement.
Moreover, this requires the medium to be transparent and the transducer's active surface to be optically accessible.
Thus, a more general approach to estimate the transducer's surface velocity is implemented by identifying a three-port Mason model using impedance measurements.

\section{Transducer modelling}

To model the electromechanical behaviour of a transducer in the spectral range around its thickness resonance frequency, the established Mason model~\cite{Mason1935} is applied.
The model uses a three-port network with two mechanical ports (with force \(F_i\) and velocity \(v_i\)) and one electrical port (with voltage \(u\) and current \(i\)).
The mechanical ports represent the faces of a given piezoelectric transducer.
The three-port can be described mathematically by a \(3 \times 3\) matrix \(\mathbf{M}\): 
\begin{equation}
	\begin{bmatrix}
		F_1 \\
		F_2 \\
		u
	\end{bmatrix}
	=
	\begin{bmatrix}
		Z\uix{m,t} \coth \left( \gamma t \right) & Z\uix{m,t} \csch \left( \gamma t \right) & h\uix{t}/(\j \omega) \\
		Z\uix{m,t} \csch \left( \gamma t \right) & Z\uix{m,t} \coth \left( \gamma t \right) & h\uix{t}/(\j \omega) \\
		h\uix{t}/(\j \omega) & h\uix{t}/(\j \omega) & 1/(\j \omega C\uix{t})\\
	\end{bmatrix}
	\cdot
	\begin{bmatrix}
		v_1 \\
		v_2 \\
		i
	\end{bmatrix}
	\label{equ:mason}
\end{equation}
with parameters
\begin{eqnarray}
	Z\uix{m,t} =& \rho\uix{t} c\uix{t} A\uix{t} , \hspace{10mm}
	\gamma &= \j \frac{\omega}{c\uix{t}} , \label{equ:mason_parameters} \\
	h\uix{t} =& c\uix{t} \sqrt{\tfrac{k\uix{t}^2 \rho\uix{t}}{\varepsilon\uix{t}}} , \hspace{10mm}
	C\uix{t} &= \tfrac{\varepsilon\uix{t} A\uix{t}}{t} . \nonumber 
\end{eqnarray}
Here, \(Z\uix{m,t}\) is the mechanical impedance of the transducer's mechanical port, defined by the product of density \(\rho\uix{t}\) and sound velocity \(c\uix{t}\) of the transducer's material and the area \(A\uix{t}\) of the transducer.
The parameter \(h\uix{t}\) couples the electrical and the mechanical properties of the model and is determined using the piezoelectric coupling factor \(k\uix{t}\) and the permittivity \(\varepsilon\uix{t}\).
Finally, the electrical capacitance of the transducer is represented by \(C\uix{t}\), which depends on the thickness \(t\) of the transducer.

\begin{figure}
	\centering
\begin{tikzpicture}[font=\footnotesize, scale=0.5]

	\draw (0,0) rectangle (4,4) node[midway] {$\mathbf{M}$};
	\draw (-2,0.5) to[short, o-] (0,0.5);
	\draw (-2,3.5) to[short, o-, i=\(v_1\)] (0,3.5);
	\draw (4,0.5) to[short, -o] (6,0.5);
	\draw (4,3.5) to[short, -o, i<=\(v_2\)] (6,3.5);
	\draw (0.5,4) to[short, -o, i<=\(i\)] (0.5,6);
	\draw (3.5,4) to[short, -o] (3.5,6);
	\draw (-2,3.5)
		to[open, v=\(F_1\)] (-2,0.5);
	\draw (6,3.5)
		to[open, v^=\(F_2\)] (6,0.5);
	\draw (0.5,6)
		to[open, v^=\(u\)] (3.5,6);

	\draw (-2,3.5) to [short] (-4,3.5)
		to [R,l_=\(Z\uix{m,1}\)] (-4,0.5)
		to [short] (-2,0.5);	

	\draw (6,3.5) to [short] (8,3.5)
		to [R,l=\(Z\uix{m,2}\)] (8,0.5)
		to [short] (6,0.5);

\end{tikzpicture}
	\caption{Three-port Mason model with the mechanical ports terminated by the mechanical impedance of the adjacent medium.}\label{fig:mason}
\end{figure}
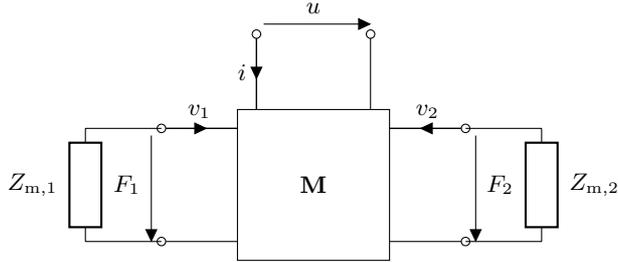
For the identification of a given transducer, the mechanical ports of the models are terminated using the mechanical impedance of the adjacent medium  (\cref{fig:mason}).
Setting \(F_i = - v_i Z\uix{m,}{}_i\), this enables to solve \cref{equ:mason} for the frequency-dependent electrical impedance \(Z\uix{el} = u / i\) of the transducer model.
In an inverse procedure, the parameters of the Mason model for a given ultrasonic transducer are identified by comparing the electrical impedance of the model \(Z\uix{el}\) with the electrical impedance \(Z\uix{meas}\) of a physical transducer~\cite{Feldmann2019}.
As the physical transducer is to be used in an acoustic absorption measurement system based on an established system for sound velocity measurement (e.g.\ applied by Javed et al.~\cite{Javed2019}), it consists of a piezoelectric disc surrounded on both sides by the target medium.
For the identification of the transducer, measurements in air are performed assuming the acoustic impedance of air (\SI{412}{\kilo\gram\per\square\meter\per\second}) multiplied by the area of the transducer to terminate the mechanical ports of the model.
For the modelling of other transducers, the specific acoustic impedance of the backing material can be used for the termination of one mechanical port with the value estimated or identified in the subsequent optimization process.
The transducer identified here consists of a hard lead zirconate titanate ceramic (PIC181, \emph{PI Ceramic GmbH}), has a thickness of \SI{0.2}{\milli\meter} and a radius of \SI{8}{\milli\meter}.
The density of the material is measured gravimetrically with the result conforming with the manufacturer's value of \SI{7800}{\kilo\gram\per\cubic\meter}.
This leaves only three parameters of the terminated Mason model to be identified: The sound velocity of the transducer's material \(c\uix{t}\), the piezoelectric coupling factor \(k\uix{t}\), and the permittivity \(\varepsilon\uix{t}\).

\begin{figure}
	\centering
	\input{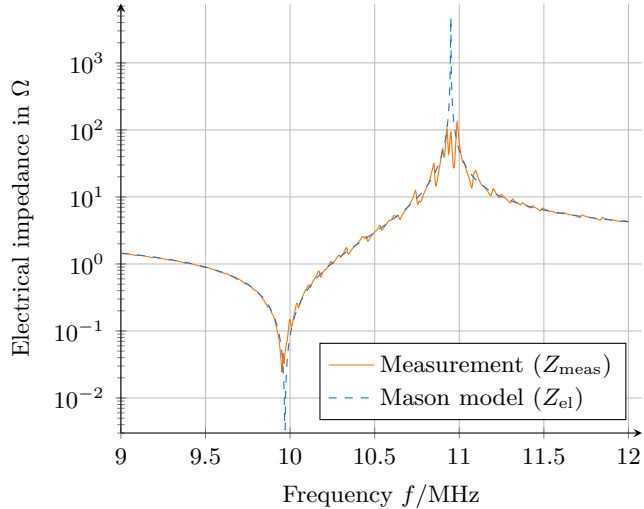}
	\caption{Magnitude of the electrical impedance of a physical transducer and the impedance of the identified Mason model.}\label{fig:mason_identified}
\end{figure}
As a cost function for the inverse procedure, the difference in the frequency-dependent magnitude of model and measured impedance is weighted with an arctangent function for robustness.
Then, the sum of the squares of this expression is minimized using a Trust Region Reflective algorithm~\cite{Branch1999}.
This optimization process yields the following parameters for the identified Mason model of the transducer: \(c\uix{t} = \SI{4380}{\meter\per\second}\), \(k\uix{t} = \SI{0.450}{}\), and \(\varepsilon\uix{t} = \SI{5.49e-9}{\ampere\second\per\volt\per\meter}\).
The resulting impedance of the model matches the measured impedance closely (\cref{fig:mason_identified}), with only the areas close to the resonance and antiresonance frequency showing significant deviation.
This is due to the low resp.\ high impedance values close to these frequencies, which result in an increased noise and uncertainty in the measurement with the impedance analyser (E4990A, \emph{Keysight Technologies}) used.
As PIC181 is a hard piezoelectric material, pronounced resonance and antiresonance frequencies are, however, expected.
The measurement also shows superimposed influence of radial modes that the Mason model cannot represent as it is based on one-dimensional considerations.

\section{Estimation of non-linearity}

\begin{figure}
	\centering
\begin{tikzpicture}[font=\footnotesize, every pin/.append style={font=\scriptsize, pin edge={black, thin}}]

\begin{axis}[
xmin=100, xmax=10000000,
xmode=log,
xmajorgrids,
xlabel={Specific acoustic impedance \(Z / \si{\kilo\gram\per\square\meter\per\second}\)},
ymin=1e-7, ymax=1e-2,
ymode=log,
ymajorgrids,
ylabel={Thermal and viscous losses \(\mu / \si{\pascal\second}\)},
axis lines=left,
enlargelimits = upper,
legend pos=south east,
legend cell align={left}
]
\addplot [color1]
table {%
100 6.15559123594945e-08
107.189131920513 6.59812164381977e-08
114.895100018731 7.07246565607341e-08
123.155060329283 7.58091031730403e-08
132.008840083142 8.12590707822428e-08
141.499129743458 8.71008361293874e-08
151.671688847092 9.33625648540946e-08
162.575566644379 1.00074447260966e-07
174.263338600965 1.07268843841279e-07
186.791359902078 1.14980441250334e-07
200.220037181558 1.23246419490987e-07
214.61411978584 1.32106631107626e-07
230.043011977292 1.4160379325241e-07
246.58110758226 1.51783693547205e-07
264.308148697411 1.62695410730667e-07
283.309610183932 1.74391551150556e-07
303.677111803546 1.86928502237026e-07
325.508859983506 2.0036670417357e-07
348.910121340677 2.14770941068941e-07
373.99373024788 2.30210653026037e-07
400.880632889846 2.46760270602864e-07
429.700470432084 2.64499573266662e-07
460.59220411451 2.8351407355557e-07
493.7047852839 3.03895428783206e-07
529.197873595844 3.25741882250801e-07
567.242606849198 3.49158736069435e-07
608.022426164942 3.74258857842101e-07
651.733960488242 4.01163223612096e-07
698.587974678525 4.30001499651315e-07
748.810385759002 4.60912665839689e-07
802.643352225717 4.94045683575918e-07
860.34644166845 5.29560211360074e-07
922.197882333432 5.67627371401124e-07
988.495904662559 6.08430570827176e-07
1059.56017927762 6.52166381313494e-07
1135.73335834311 6.99045481193213e-07
1217.38272773966 7.49293664378008e-07
1304.9019780144 8.03152920690463e-07
1398.71310264724 8.60882592495904e-07
1499.26843278605 9.22760612817837e-07
1607.05281826164 9.89084830426604e-07
1722.58596539879 1.06017442770288e-06
1846.42494289554 1.13637143739364e-06
1979.16686785356 1.21804236469386e-06
2121.45178491063 1.30557992139769e-06
2273.96575235793 1.399404879161e-06
2437.44415012222 1.49996804919467e-06
2612.67522556333 1.60775239595351e-06
2800.50389418363 1.72327529258564e-06
3001.83581357559 1.84709092604297e-06
3217.64175025074 1.97979285980455e-06
3448.96226040576 2.12201676210135e-06
3696.91270719502 2.27444330732144e-06
3962.68863870148 2.43780125787361e-06
4247.5715525369 2.6128707331447e-06
4552.93507486695 2.80048667123472e-06
4880.25158365443 3.00154248781581e-06
5231.09930805626 3.216993934637e-06
5607.16993820546 3.44786315776913e-06
6010.27678207038 3.69524295250541e-06
6442.36350872137 3.96030120772847e-06
6905.51352016233 4.24428552730942e-06
7401.95999691564 4.54852800946228e-06
7934.09666579749 4.87445015662913e-06
8504.48934180268 5.22502261563752e-06
9115.88829975082 5.60363897693138e-06
9771.2415353465 6.00980402841019e-06
10473.7089795945 6.44552069866707e-06
11226.6777351081 6.91293308945557e-06
12033.7784077759 7.41433479821341e-06
12898.9026125331 7.95217725365602e-06
13826.2217376465 8.52907791220953e-06
14820.2070579886 9.14782811708255e-06
15885.6512942805 9.81140036580914e-06
17027.691722259 1.0522954664632e-05
18251.8349431904 1.12858435676795e-05
19563.9834351706 1.21036154042704e-05
20970.4640132323 1.29800150880535e-05
22478.0583354873 1.39189817770627e-05
24094.0356023953 1.49246425154954e-05
25826.1876068267 1.60013008394941e-05
27682.8663039206 1.71534191768404e-05
29673.0240818887 1.84002725908742e-05
31806.2569279412 1.97707054406594e-05
34092.8506974681 2.12420229680352e-05
36543.8307095725 2.28207894012337e-05
39171.0149080926 2.45137304220642e-05
41987.0708444391 2.63276661885019e-05
45005.576757005 2.82694268331919e-05
48241.0870416537 3.03457492944749e-05
51709.2024289676 3.25631553069973e-05
55426.645206631 3.49278118471528e-05
59411.3398496504 3.74711691045157e-05
63682.4994471859 4.03280508542594e-05
68260.7183427238 4.33824299720884e-05
73168.0714342719 4.66417984958291e-05
78428.2206133768 5.01126942981354e-05
84066.5288561832 5.38005411206859e-05
90110.1825166502 5.77095392261913e-05
96588.3224115871 6.18426313853438e-05
103532.184329566 6.62015685202073e-05
110975.249641207 7.09040600964143e-05
118953.406737032 7.62646933491115e-05
127505.124071301 8.19333102577192e-05
136671.6356462 8.79135933448196e-05
146497.139830728 9.42092284492421e-05
157029.012472938 0.000100824499459773
168318.035333095 0.000107764900324064
180418.640939207 0.000115037694234517
193389.175045523 0.000122652346323065
207292.177959537 0.000130620762917332
222194.686093952 0.000138957287140093
238168.555197616 0.000147678425071822
255290.806823952 0.000156962542517717
273643.999707467 0.000166950651536314
293316.627839004 0.00017736218310825
314403.54715915 0.000188206594993132
337006.432927192 0.000199491690221422
361234.269970943 0.000211222678818653
387203.878181256 0.000223401234672113
415040.475785047 0.000236024576227953
444878.283112759 0.00024908458887689
476861.169771447 0.000262972318143289
511143.348344017 0.000277947943193114
547890.117959393 0.000293483464818291
587278.661318948 0.000309563965421255
629498.899022189 0.000326167017975468
674754.405311069 0.000343262068680821
723263.389648353 0.00036080988551293
775259.748862946 0.000378762084055279
830994.194935339 0.000397060742587218
890735.463861044 0.000415638121019427
954771.611420807 0.000434416503573678
1023411.40210545 0.000453308192539343
1096985.79789238 0.000472215688350363
1175849.55405216 0.000491032097421975
1260382.92967973 0.000509641811664742
1350993.52119803 0.000528346094329861
1448118.22767453 0.000550336589578388
1552225.35742705 0.000572251695530369
1663816.88607613 0.000593970699248269
1783430.87693191 0.000615366723408081
1911644.0753857 0.000636308479803712
2049074.68981585 0.000656662361903711
2196385.37241655 0.000676294845480035
2354286.41432242 0.000695075145567356
2523539.17043477 0.000712878056552201
2704959.73046313 0.000729586881519663
2899422.85388288 0.000745096338977148
3107866.18778201 0.000759315321808611
3331294.78793467 0.000772169376828878
3570785.96490046 0.0007860368294323
3827494.47851631 0.000803881232201147
4102658.10582719 0.000820594155192095
4397603.60930272 0.000836119190490912
4713753.13411672 0.000850416618819246
5052631.06533568 0.000863463965139393
5415871.37807946 0.000875255998679427
5805225.5160949 0.000885804181623858
6222570.83673023 0.000895135602735884
6669919.66303011 0.000903291461447159
7149428.98659758 0.000910325191799643
7663410.86800745 0.000916300331981695
8214343.58491942 0.000921288252867095
8804883.58164346 0.000925365857747609
9437878.27777537 0.000928613356139037
10116379.7976621 0.000931112198745893
10843659.6868961 0.000932943240535611
11623224.6867985 0.000934185176747612
12458833.6429501 0.000934913274772902
13354515.629299 0.000935198405032906
14314589.3752348 0.000939862086490788
15343684.0893001 0.000944368780249343
16446761.7799466 0.000948373141160509
17629141.1809595 0.000951923088930321
18896523.3969121 0.000955063717307778
20255019.3923067 0.000957837061928378
21711179.456945 0.000960281983648695
23272024.7896041 0.000962434146455622
24945081.3523032 0.000964326070008268
26738416.1583994 0.000965987238648021
28660676.1694825 0.000967444250952178
30721129.9886175 0.000968720996334968
32929712.5509715 0.000969838847613629
35297073.0273065 0.000970816860718128
37834626.1713193 0.00097167197474816
40554607.3584083 0.000972419207330113
43470131.5812502 0.000973071841691151
46595256.6866468 0.000973641603059789
49945051.1585514 0.000974138822947482
53535666.7741072 0.000974572590595102
57384416.4830239 0.000974950891415205
61509857.885805 0.000975280732657786
65931882.7133354 0.000975568256803405
70671812.7392749 0.000975818843368895
75752502.5877191 0.000976037199919346
81198449.9318401 0.000976227443133778
87035913.6148517 0.000976393170785659
93293040.2628468 0.000976537525484705
100000000 0.000976663250992378
};
\addlegendentry{\(\mu =  \frac{v_0 Z}{\omega}\)}
\addplot [only marks, color2]
table{%
x                      y
1479270.3343691 0.00134130866757647
915088.334220801 0.00160971039804502
883492.721427708 0.000798676897016101
523.249632329355 5.21816442197669e-05
165.473655965641 4.58427233199355e-05
950.849185297312 5.32054954625257e-05
408.278241869063 3.45963986021279e-05
711155.04215864 0.000433066984510731
};

\node[pin={45:{Water}}] at (axis cs: 1479270.3343691, 0.00134130866757647){};
\node[pin={90:{Ethanol}}] at (axis cs: 915088.334220801, 0.00160971039804502){};
\node[pin={180:{Methanol}}] at (axis cs: 883492.721427708, 0.000798676897016101){};
\node[pin={-45:{Argon}}] at (axis cs: 523.249632329355, 5.21816442197669e-05){};
\node[pin={45:{Helium}}] at (axis cs: 165.473655965641, 4.58427233199355e-05){};
\node[pin={0:{Xenon}}] at (axis cs: 950.849185297312, 5.32054954625257e-05){};
\node[pin={-90:{Air}}] at (axis cs: 408.278241869063, 3.45963986021279e-05){};
\node[pin={-90:{n-Hexane}}] at (axis cs: 711155.04215864, 0.000433066984510731){};

\end{axis}

\end{tikzpicture}
	\caption{Minimum value for losses \(\mu \) dependent on the specific acoustic impedance of the target medium.}\label{fig:losses_nonlinearity}
\end{figure}
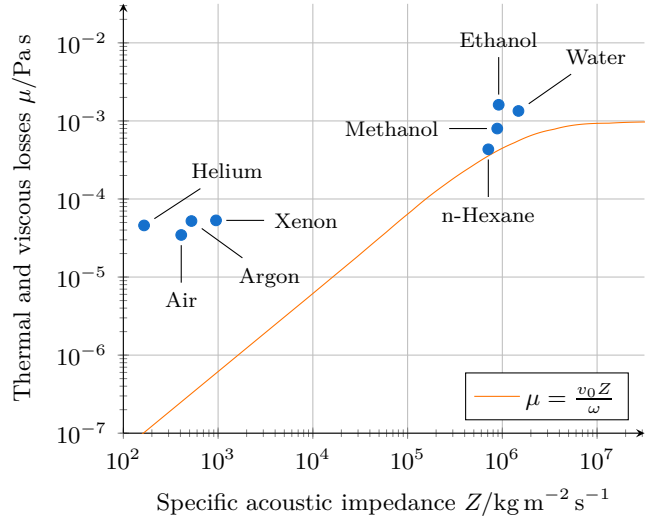

With the Mason model identified in the previous section, it is possible to model the electromechanical behaviour of the transducer.
This allows to estimate the velocity of the faces of the transducer for a given voltage.
Changing the terminating mechanical impedance also allows to estimate the velocity for changing target media.
Assuming that the transducer is terminated with the same mechanical impedance (\(Z\uix{m} = A\uix{t} Z\)) at both mechanical ports as before, solving \cref{equ:mason} for \(v / u\) yields the frequency response of the transducer in transmission mode:
\begin{equation}
	\frac{v}{u} = G\uix{t}(\j\omega) = {\left(\frac{Z\uix{m}}{C\uix{t} h\uix{t}} - \frac{Z\uix{m,t}}{C\uix{t} h\uix{t}} \left(\coth \left( \gamma t \right) + \csch \left( \gamma t \right)\right) - \frac{2 h\uix{t}}{\j \omega} \right)}^{-1} .
	\label{equ:frequency_response}
\end{equation}
Note that \(\gamma \) also depends on the angular frequency \(\omega \) (\cref{equ:mason_parameters}). 
For continuous, monofrequent excitation of acoustic waves, one can apply \cref{equ:frequency_response} directly by setting \(u\) to the electrical signal's amplitude and solving for the velocity \(v\).
The absolute value of \(v\) can then be used as an estimate for the particle velocity close to the transducers surface \(v_0\), allowing to determine the acoustic Reynolds number \(N\uix{Re}\) using \cref{equ:acoustic_reynolds}.
In physical measurement systems, however, signals are typically limited in the temporal regime and thus have a finitely small bandwidth.
As \cref{equ:frequency_response} models the frequency-dependent behaviour of the transducer, it describes how an electrical voltage signal translates into a velocity signal in the frequency domain. 
Fourier transform (\(\mathcal{F} \left \{\right \} \)) and inverse Fourier transform (\(\mathcal{F}^{-1} \left \{\right \} \)) then allow to model the influence of the transducer on a transient signal \(u(t)\):
\begin{equation}
	v(t) = \mathcal{F}^{-1} \left \{ G\uix{t}(\j\omega) \mathcal{F} \left \{ u(t) \right \} \right \} .
\end{equation}
The resulting transient velocity \(v(t)\) of the transducer's faces can then be evaluated for its maximum as an estimate for \(v_0\):
\begin{equation}
	v_0 \approx \max(v(t)) .
\end{equation}
Similar to the physical setup~\cite{Javed2019}, the electrical excitation signal \(u(t)\) is modelled as a Gaussian modulated sinusoidal pulse with a centre frequency of \SI{10.5}{\mega\hertz}, a relative bandwidth of \SI{0.1}{}, and a peak voltage of \SI{1}{\volt}.
As the setup utilizes the pulse-echo technique, the centre frequency is chosen between the resonance and antiresonance frequency (\cref{fig:mason_identified}) of the transducer to enable transmitting and receiving operation.
The setup is used for a variety of different fluids, so the maximum of the velocity \(v_0\) is evaluated dependent on the specific acoustic impedance \(Z\) of the fluid.
The results are then inserted in \cref{equ:losses_bound} to determine the lower bound of the linear losses \(\mu \) the fluid to be analysed needs to have for the acoustic Reynolds number to be less than one, resulting in predominately linear wave propagation.
To analyse a wide range of the specific acoustic impedance, the results are presented with logarithmic scales (\cref{fig:losses_nonlinearity}), showing that the  minimal losses \(\mu \) for linear sound propagation increase with the specific acoustic impedance \(Z\) of the target fluid.	
At values for the specific acoustic impedance of the fluid that approach and exceed the specific acoustic impedance of the transducer's material, the minimal losses for linear sound propagation show a constant value. 
Note that these results are only valid for the setup and transducer described before with an excitation signal voltage of \SI{1}{\volt}.

As a reference, the specific acoustic impedances and losses of several fluids at \SI{293}{\kelvin} and \SI{100}{\kilo\pascal} are included in \cref{fig:losses_nonlinearity} as well~\cite{Lemmon2013}.
The losses \(\mu \) depicted are low estimates, as they only include the influence of shear viscosity and thermal conductivity (\(\mu = \frac{4}{3} \mu\uix{s} + \frac{c\uix{p} - c\uix{v}}{c\uix{p} \cdot c\uix{v}} \nu \)), omitting the additional loss caused by the relatively unexplored volume viscosity.
Thus, in a physical setup, the difference between the minimal losses necessary for linear sound propagation and the actual losses in the respective fluid is expected to be more significant.

The sound propagation in all fluids used for the comparison is expected to be predominantly linear, if the identified transducer and the excitation signal is used.
The distance to the lower boundary for the losses is significant for the gases used for comparison (helium, air, argon and xenon), showing that the acoustic Reynolds number of these transducer-fluid combinations is significantly smaller than one.
For the depicted liquids (n-hexane, methanol, ethanol and water), the distance to the depicted graph is smaller.
Especially if n-Hexane is analysed with the identified transducer, the acoustic Reynolds number approaches one.
In this case, non-linear wave propagation may occur and measures to prevent or detect these non-linear effects, such as lowering the signal voltage or analysing the acoustic signal spectrum for higher harmonics, should be taken.
It should be noted that changing the thermodynamic state of the fluids will result in different properties (\(Z\) and \(\mu \)) which could potentially fail to satisfy \cref{equ:losses_bound}.
Also note that these considerations constitute a worst-case assessment, as additional dissipative effects that may prevent non-linear wave propagation in the fluid, such as the effects of losses in the transducer's material and additional linear absorption due to volume viscosity, are neglected.

\section{Conclusions}

A means to assess whether acoustic wave propagation can be assumed as linear in a given medium is the acoustic Reynolds number.
The evaluation of this parameter, however, requires quantitative information about the particle velocity.
This velocity can be estimated using a Mason model for a given transducer, which can be identified by a measurement of the transducer's frequency-dependent electrical impedance.
The procedure requires no direct measurement of the particle velocity or other acoustic quantities and is thus easy to realize experimentally for a variety of application scenarios.

The approach may be further expanded by applying more in-depth models for the transducers, using e.g.\ chain matrices for the modelling of matching layers~\cite{Webersen2105} or complete finite-element simulations.
As the results of these consideration describe a worst-case scenario (if the aim is to have linear sound propagation), edge cases (\(N\uix{Re} \approx 1\)) should be reviewed by evaluating the acoustic signal spectrum for the existence of harmonic frequencies caused by non-linearity. 

\printbibliography[]

@article{Branch1999,
	doi = {10.1137/s1064827595289108},
	year = {1999},
	month = 1,
	publisher = {Society for Industrial {\&} Applied Mathematics ({SIAM})},
	volume = {21},
	number = {1},
	pages = {1--23},
	author = {Mary Ann Branch and Thomas F. Coleman and Yuying Li},
	title = {A Subspace,  Interior,  and Conjugate Gradient Method for Large-Scale Bound-Constrained Minimization Problems},
	journal = {{SIAM} Journal on Scientific Computing}
}

@article{Javed2019,
	doi = {10.1021/acs.jced.8b00938},
	year = {2019},
	month = 2,
	publisher = {American Chemical Society ({ACS})},
	volume = {64},
	number = {3},
	pages = {1035--1044},
	author = {Muhammad Ali Javed and Elmar Baumh\"{o}gger and Jadran Vrabec},
	title = {Thermodynamic Speed of Sound Data for Liquid and Supercritical Alcohols},
	journal = {Journal of Chemical {\&} Engineering Data}
}

@article{Feldmann2019,
	author = {Feldmann, Nadine and Jurgelucks, Benjamin and Claes, Leander and Schulze, Veronika and Henning, Bernd and Walther, Andrea},
	year = {2019},
	title = {An inverse approach to the characterisation of material parameters of piezoelectric discs with triple-ring-electrodes},
	pages = {59--65},
	volume = {86},
	number = {2},
	issn = {0171-8096},
	journal = {tm - Technisches Messen},
	doi = {10.1515/teme-2018-0066}
}

@book{Landau2013,
	title = {Fluid Mechanics},
	author = {Landau, L.D. and Lifshitz, E.M.},
	number = {v. 6},
	isbn = {978-1-4831-4050-6},
	year = {2013},
	publisher = {Elsevier Science}
}

@article{Lemmon2013,
	title = {REFPROP 9.1},
	author = {Lemmon, E. W. and Huber, M. L. and McLinden, M. O.},
	journal = {NIST Standard Reference Database},
	volume = {23},
	year = {2013}
}

@article{Mason1935,
	author = {Mason, W. P.},
	title = {An Electromechanical Representation of a Piezoelectric Crystal Used as a Transducer},
	journal = {Bell System Technical Journal},
	volume = {14},
	number = {4},
	pages = {718-723},
	year = {1935},
	month = {10},
	doi = {10.1002/j.1538-7305.1935.tb00713.x}
}

@book{Rudenko1977,
	author = {Rudenko, Oleg Vladimirovi{\v{c}} and Solujan, Stepan Ivanovich and Beyer, Robert T.},
	year = {1977},
	title = {Theoretical foundations of nonlinear acoustics},
	address = {New York},
	publisher = {{Consultants Bureau}},
	isbn = {978-1-4899-4796-3},
	series = {Studies in Soviet science Physical sciences}
}

@inproceedings{Webersen2105,
	doi = {10.5162/SENSOR2015/B1.2},
	author = {M. Webersen and F.  Bause and J. Rautenberg and B. Henning},
	title = {B1.2 - Identification of Temperature-Dependent Model Parameters of Ultrasonic Piezo-Composite Transducers},
	publisher = {AMA Service GmbH, Germany},
	year = {2015},
	booktitle = {AMA Conferences 2015}
}

\end{document}